\documentclass[aps,prl,twocolumn,superscriptaddress,nofootinbib,
nobalancelastpage,showpacs,nobibnotes]{revtex4}
\usepackage{epsfig}

\begin{document}
\title
{Electron capture rates in a plasma
}

\author{R. F. Sawyer}
\affiliation{Department of Physics, University of California at
Santa Barbara, Santa Barbara, California 93106}

\begin{abstract}

A new general expression is derived for nuclear electron capture rates within dense plasmas.
Its qualitative nature leads us to question some widely accepted assumptions about how to calculate
the effects of the plasma on the rates.
A perturbative evaluation, though not directly applicable to the 
strongly interacting case, appears to bear out these suspicions.

\end{abstract}
\maketitle

\section{Introduction}

Calculation of electron capture rates in plasmas is of importance in many astrophysical systems. 
The influence of the surrounding plasma on the capture rates is completely understood for the case of
a weakly coupled plasma in which the electrons are not degenerate, as long as the energy release
in the reaction, $Q$, is sufficiently large.  These conditions are well fulfilled in the solar core, where the capture rate of electrons on $^7$Be has been studied in detail. Here the weak-screening Salpeter factor \cite{salp}, of a multiplicative factor of $\exp(-e^2Z \kappa_D/T)\approx (1-e^2 Z \kappa_D/T)$ gives a 16\%  rate reduction \cite{iben}, \cite{bg},\cite{bs2}. \footnote{The question of whether or not the exponentiation of the perturbative result is justified is numerically irrelevant in this case.} These calculations of reduction factors are all really just calculations
of a screening-induced reduction of the expectation of the electron density at the position of the ion.
The methods of the above references are not applicable to cases with: a) strong plasma coupling $\Gamma$; or b),  small energy release $Q$ in relation to other energies in the problem
(or negative $Q$). Indeed, as we explicitly show below, the plasma-induced 
change of the electron density at the position of the ion is no longer the determining factor
in the rate when condition b) prevails. Also, of the approaches in the above references,
only that of ref. \cite{bs2} can incorporate high electron degeneracy\footnote{ although we know of
no very interesting set of conditions that are highly degenerate without also embodying condition
a) or condition b).}

In the parameter region relevant to explosions/collapse of O-Ne-Mg stellar cores, for
example,
both conditions a) and b) apply and the electrons are very degenerate.
Indeed $Q$ is here negative for important capture reactions; they become enabled when the Fermi energy
increases nearly to the value $-Q$. In ref. \cite{gut}, addressing these systems, we find a simple statement of a procedure that appears to be widely followed \cite{loumas}, \cite{itoh}, \cite{bravo}, \cite{lang} ; namely, to calculate the
capture rate for an individual electron as in vacuum, except replacing the vacuum $Q$ value for the decay by an effective value
$Q+E_1-E_2$ where $E_1$ and $E_2$ are the respective energies of interaction of the initial and final ions in the plasma.
There exist more or less standard expressions for these energies derived from classical simulations.
For the purpose of our present discussion we designate this correction, with energies calculated as for
a classical plasma, as the ``ionic energy correction." In applications, it appears that this correction has been applied
sometimes with accompanying initial state Coulomb effects involving the electrons, which we loosely characterize as ``screening", and sometimes without.

The present work begins with the derivation of a new expression for the rate of electron capture, one that we think
should be the basis of future calculation of rates in the presence either of strong degeneracy, or of strong
plasma coupling. Just looking at the form of the result, we believe, should be enough to make one profoundly skeptical
about assumptions being made in the literature. We then carry out a systematic perturbative evaluation in the weakly
coupled case, for cases of arbitrary electron degeneracy, retaining terms of relative order 
 $e^2 Z \kappa_D/T$ in the rates (or of order ``Salpeter" as we shall henceforth designate them). The results appear to be
 in complete disagreement with the assumptions
 that lead to the ionic energy correction, had they been applied to a weakly coupled problem. 

We use the notation $\kappa_D$ for the screening wave number throughout. In the presence of 
some degree of degeneracy its square is given by \cite{bs},

\begin{eqnarray}
\kappa_D^2=4\pi \Bigr [\beta \sum_s  e_s ^2 n_s  +e^2 {\partial \over \partial \mu_e}\int {d{\bf p}\over (2 \pi)^3} f(E_p)\Bigr],
\label{debye}
\end{eqnarray}
where $e_s=e Z_s$ is the charge of ionic species $s$, $n_s$ the number density of
that species, and $f(E_p)$ the Fermi distribution. In the non-degenerate limit this gives the Debye wave
number, and in the strongly degenerate limit, the electronic part would give the square of the
Thomas-Fermi screening wave number.

\section{ General formulation}
Our general methods follow those introduced in ref. \cite{bs}.
We take the electron capture interaction Hamiltonian for the process of a non-relativistic electron plus ion, $I_a$, going into neutrino plus ion $I_b$ to be given by a zero range
Fermi type coupling (all the effects of Coulomb interactions in what follows being the same for the Gamow-Teller case), 
\begin{equation}
H_W=\int d {\bf r}[ K({\bf r},t)+K^\dagger({\bf r},t)] ,
\label{hw}
\end{equation}
where
\begin{eqnarray}
K({\bf r},t)=ge^{-i Q t} \psi_{\nu}^\dagger ({\bf r},t) \psi_e ({\bf r},t)\psi_b^\dagger ({\bf r},t) 
\psi_a ({\bf r},t),
\label{K}
\end{eqnarray}
and $Q$ is the energy release in the reaction. There are suppressed spin indices contracted with each other on the electron and neutrino fields.
Here the $\psi_{\{a,b\}}$ are nonrelativistic quantum fields that describe creation or annihilation of
the respective ions. All fields are in a Heisenberg picture with respect to the complete Hamiltonian. The ionic fields could be Fermi or Bose, but we shall remain in a domain of temperature and density in which
deviation from Boltzmann statistics of the ions is inconsequential.

We are interested in the time rate of change of the electron density induced by (\ref{hw}). Since the medium is translationally invariant we can choose to evaluate this time derivative at 
at point ${\bf r}={\bf 0}$ and choose time to be zero as well. Directly from the Heisenberg equations we obtain
the rate of change, $w$, of the electron density at the origin, $n_e({\bf 0},0)=\psi_e^\dagger({\bf 0},0)\psi_e ({\bf 0},0)$,
\begin{eqnarray}
w=\Bigr \langle \dot n_e({\bf 0,0})\Bigr \rangle_\beta=-i\int d{\bf r}\Bigr \langle [n_e({\bf 0},0),K({\bf r}, 0)]\Bigr \rangle_\beta
\nonumber\\
=i\Bigr \langle[K({\bf 0},0)-K^\dagger({\bf 0},0)]\Bigr \rangle_\beta.
\label{ndot}
\end{eqnarray}

The notation $\langle...\rangle_\beta$ indicates the thermal average in the medium, such that for an operator, $O$, we have
 $\langle O\rangle_\beta\equiv Z_P^{-1}{\rm Tr} [O\exp(-\beta [H+H_{W}- \mu_e N_e])]$ where $Z_P$ the partition function, and where $H+H_W$ is the complete Hamiltonian. We wish to calculate the rate to
lowest non-vanishing order; i.e. to second order in the weak coupling parameter $g$. So it is it clear that one power of g must come from the weak interaction
$H_{W}$ within the statistical factor. Thus we now must consider the linear response \cite{fw} of the medium average of the operator $[K(0)-K^\dagger(0)]$ to this perturbation,
\begin{eqnarray}
&w=-i \int_{-\infty}^0 dt \int (d{\bf r})\Bigr \langle [\dot n_e({\bf }0,0),H_W({\bf r}]\Bigr \rangle_\beta=-\int_{-\infty}^0 dt
\nonumber\\
&\times \int d{\bf r}\Bigr \langle [K({\bf 0},0)-K^\dagger({\bf 0},0)],[K({\bf r},t)+K^\dagger({\bf r},t)]\Bigr \rangle_\beta ,
\nonumber\\
& \,
\end{eqnarray}
where now the thermal average in the medium is to be calculated using $H$ alone. $H$ conserves electron
number, whence,

\begin{eqnarray}
\Bigr\langle [K({\bf 0},0), K({\bf r},t)]\Bigr \rangle_\beta=
\Bigr\langle [K^\dagger ({\bf 0},0), K^\dagger({\bf r},t)]\Bigr \rangle_\beta=0.
\label{Kcom}
\end{eqnarray}
Using in addition the space-time translational invariance of the medium and the antisymmetry of
the commutator,
\begin{eqnarray}
\Bigr\langle [K^\dagger ({\bf 0},0), K({\bf r},t)]\Bigr \rangle_\beta=-\Bigr\langle [K ({\bf 0},0), K^\dagger({\bf -r},-t)]\Bigr \rangle_\beta ,
\end{eqnarray}
we can write the rate as
\begin{eqnarray}
w
=-\int_{-\infty}^\infty dt \int d{\bf r}\Bigr \langle [K({\bf 0},0),K^\dagger({\bf r},t)]\Bigr \rangle_\beta 
\label{w2} \,.
\end{eqnarray}

When we take the medium to contain no ions of type $b$, so that there is no reverse reaction, we can omit the
first term in the commutator in (\ref{w2}), which, after inserting (\ref{K}), gives,
\begin{eqnarray}
 w=g^2  \int_{-\infty}^\infty dt \int d{\bf r}\,e^{-i Q t}
\Bigr\langle \psi_e^\dagger({\bf r},t)\psi_a^\dagger({\bf r},t)
\nonumber\\
 \times \psi_{\nu}({\bf r},t)\psi_b({\bf r},t) 
 \psi_b^\dagger({\bf 0},0)\psi_{\nu}^\dagger({\bf 0},0)\psi_a({\bf 0},0)\psi_e({\bf 0},0)
\Bigr \rangle_\beta
\nonumber\\
=g^2 \int_{-\infty}^\infty dt \int d{\bf r}\,e^{i Q t}\int {d^3 p_{\nu}\over (2\pi )^6} e^{i {\bf p_\nu\cdot r}+iE_\nu t}
\nonumber\\
\times \langle \psi_e^\dagger({\bf r},t)\psi_a^\dagger({\bf r},t)
 \psi_b({\bf r},t) 
 \psi_b^\dagger({\bf 0},0)\psi_a({\bf 0},0)\psi_e({\bf 0},0)\Bigr\rangle_\beta.
\nonumber\\
\,
\label{orig}
\end{eqnarray}

For almost any purposes the thermal motions and recoil of the reacting ions are ignorable. We distinguish the ion
$I_a$ on which an electron is captured and the capture product $I_b$ from the other ions in the plasma, taking a
single $I_a$  fixed at the origin, replacing ${\bf r}={\bf 0}$ in all of the explicit fields in (\ref{orig}) and adding an all-over factor of density of ions of type $a$, $ n_a$. The products of the field operators $\psi_{a,b}$
that occur in (\ref{orig}) now serve only to change one ion into the other. We substitute
\begin{eqnarray}
\psi^\dagger_a({\bf 0},t)\psi_b({\bf 0},t)=e^{iH t}|a\rangle\langle b |\,e^{-iHt},
\end{eqnarray}
where the time dependence reflects the fact that the plasma interactions of the two ions are
different. We obtain,
\begin{eqnarray}
w={g^2  n_a\over 2 \pi^2}  
\int_{-\infty}^\infty dt \,e^{-i Q t}\int dp_\nu \,p_{\nu} e^{iE_\nu t}M(t),
\label{orig2}
\end{eqnarray}
where
\begin{eqnarray}
M(t)=Z_P^{-1}\times~~~~~~~~~~~~~~~~~~~~~~~~~~~~~~~~
\nonumber\\
{\rm Tr}\Bigr [\langle a| ^{-\beta H }e^{i H t} \psi_e^\dagger ({\bf 0},0) |a \rangle \langle b |e^{-i H t}| b \rangle  \langle a|\psi_e({\bf 0},0)|a \rangle \Bigr ].
\nonumber\\
\,
 \label{M1}
\end{eqnarray}
The ionic states $| a\rangle  $,$| b\rangle $, in (\ref{M1}) serve to determine the coupling
to the fixed ion in the interaction Hamiltonian. The trace in (\ref{M1}) is over all of the other coordinates. We write
\begin{eqnarray}
M(t)=Z_P^{-1}{\rm Tr}\Bigr [ e^{-(\beta-it) H_a }\psi_e^\dagger ({\bf 0},0)  e^{-i H_b t}\psi_e({\bf 0},0)
\Bigr ],
\nonumber\\
\,
 \label{M2}
\end{eqnarray}
where
\begin{eqnarray}
Z_P={\rm Tr}\Bigr[ e^{-\beta H_a+\beta \mu_e N_e} \Bigr ]\,.
\end{eqnarray}
Here $H_a$ is the Hamiltonian for the initial system (with $H_W$ turned off, since our expressions are already of
order $g^2$) and $H_b$ that for the final system, not counting the constant $Q$. The space of states in which
the trace in (\ref{M2}) is to be evaluated is just what we have defined as the ``plasma"; all electrons, and all ions
except the ions $a$ and $b$. To restate the definitions of $H_a$ and $H_b$, they are the Hamiltonians for the
complete plasma in the presence of the respective fixed Coulomb fields of the nuclei, $a,b$.

The result (\ref{M2}) for the function $M(t)$, in conjunction with the rate expression (\ref{orig2})
is our fundamental result. We show the details of the perturbation evaluation in the appendices, retaining
terms of order Salpeter ($e^2 Z\kappa_D$). This calculation appears to be
very inefficient, given the simplicity of the result, but we have found no other path
to the answer. We state the results in terms of two subsidiary functions, $M_0(t)$ and $M_A(t)$, 
\begin{eqnarray}
M_0(t)= \int {d{\bf p}_1\,d{\bf p}\over (2 \pi)^3}
\Bigr \langle a_{{\bf p}_1}^\dagger a_{\bf p} e^{i E_{\bf p}t} \Bigr \rangle
= \int {{d^3p\over (2\pi)^3} }f(E_p)e^{iE_pt},
\nonumber\\
\,
\label{m0}
\end{eqnarray}
and
\begin{eqnarray}
M_A(t)= \int {d^3 p \over (2 \pi)^3}[f(E_p)]^2 e^{iE_pt},
\label{mA}
\end{eqnarray}
where $f(p)$ is the Fermi distribution,
\begin{eqnarray}
f(E_p)=[1+e^{\beta(E_p-\mu_e)}]^{-1}.
\label{fermi}
\end{eqnarray}
The kernel of the rate expression (\ref{orig2}) is now given by,
\begin{eqnarray}
 M(t)=M_0(t)-e \,e_a  \kappa_D \beta\Bigr (M_0(t)-M_A(t)\Bigr ),
\label{ans4}
\end{eqnarray}
and the rate by,
\begin{eqnarray}
w=2^{3/2} \pi g^2  m^{3/2}\int_{{\rm min}(0,-Q)}^\infty d E\, E^{1/2}\, (Q+E)
\nonumber\\
\times \Bigr[ f(E) -e e_a   \kappa_D \beta 
[f(E)-f^2(E)]\Bigr].
\label{finalans}
\end{eqnarray}
We note that to the order that we are working, the result  (\ref{finalans}) agrees exactly with
eq. 1.21 of ref. \cite{bs2} except for the additional term, $E$ in the factor $Q+E$ in our
result, and the inclusion of all vacuum Coulomb effects in ref. \cite{bs2}. ( In the present work 
we just stated that vacuum Coulomb effects could be treated additively.) In the vacuum case
the extra $E$ term just puts in corrected kinematics. But, as we see more explicitly in the next
section, once we make this trivial modification, the rate is no longer determined by
the electron density at the location of the ion. Refs. \cite{bg} and \cite{bs2} calculate
only the perturbation of the density.
Thus it is of some interest that this modification is screened by the same factor (under the $E$ integral) as the $Q$ term, but it is perhaps not unexpected. 
What is more of note is not what our result contains, but what it does not contain. We discuss
this at some length in the next section.

.
\section{Discussion}

We return
to the basic result for the kernel of the capture rate formula,  

\begin{eqnarray}
M(t)=Z_P^{-1}Tr \Bigr [e^{\beta \mu_e N_e} e^{(-\beta +i t)H_a }\psi_e^\dagger ({\bf 0},0)  e^{-i H_b t}\psi_e({\bf 0},0)
\Bigr ].
\nonumber\\
\,
 \label{M3}
\end{eqnarray}
If we were to replace both factors $\exp(i H_a t)$ and $\exp(-i H_b t)$ by unity, i. e. evaluate
at $t=0$, then $M$ would be precisely the electron density at the position of the ion. Then, in accord with
the remarks at the end of the last section, we get the standard screening
results, where $Q+E$ in (\ref{finalans}) is replaced by $Q$. In our mechanics this
replacement comes from the fact that Hamiltonians, $H_a$ and
$H_b$ both contain electron kinetic energy terms, and that there is an electron creation operator
standing between the factors containing them; thence producing an $\exp [-i E t]$. But $H_a$ and $H_b$ also contain pieces in which the initial and final ions interact with the plasma. So in parallel fashion, why do we not see another addition to $Q$
in the results, namely $\Delta Q=\Delta E_a-\Delta E_b$, due to the ionic interaction shifts in
the plasma? After all, one can infer
energy shifts for the ions at the Salpeter level. For an ion with charge $e_a$ we have $\Delta E_a=e_a^2 \kappa_D/2$
so that the difference between the energy shifts of the initial and final ions in our case would be
$e e_a \kappa_D$. But there is no such term in the result (\ref{finalans}).

This is our first criticism of the ``ionic energy shift" ansatz. It should have manifested itself
at our level of approximation and it did not. A reader might comment, ``I look at (\ref{finalans}), say in the non-degenerate limit, and
I see just Salpeter screening, an effect that pertains only to physics in the initial state. But we know that
the energies of these ions are shifted differently by the plasma interactions, and when these shifts
are of the order Q, they must matter a lot. So the author cannot have put in all of the physics."
But if one wades through the calculations in the appendix, one finds that such terms come and
go; they all cancel in the end. What is left out of the ansatz, [initial state effects + final state effects],
is the fact that many of the individual terms in the development, even to this order, involve one Coulomb
interaction with the initial ion and one with the final one. These terms cancel almost all of what comes from the ansatz.
We see no valid reason to believe that the ionic energy shift correction by itself, as it is applied
in refs. \cite{loumas}, \cite{itoh}, \cite{bravo}, \cite{lang} will serve us any better in the strongly coupled case than in our case in which we can really calculate.

 In addition
to this ionic energy shift, there exists a second plausible correction recommended in some papers dealing with
extremely dense and degenerate systems \cite{itoh},\cite{lang}, namely an electron energy shift
resulting from the screening of the potential seen by the electron. The qualitative description
of how this screening should affect the rates is now nothing like the description in weak screening
theory; in the latter case the effects are entirely due to the change in electron density at the
ionic position and are not directly connected to the energy spectrum of the electrons.
But when we address a degenerate case in which $Q<0$ and we get captures 
only as the Fermi energy approaches $-Q$, it is the energy distribution rather than the electron density 
that would
seem to be the critical feature.

To discuss this suggested effect, we first note that in the derivation of our perturbative results,
given in the appendix,
we did need to calculate  a change in the electron chemical potential, $\delta \mu_e$,
Our primary calculation was for fixed chemical potential and we thereafter made a
correction to get the results for fixed electron density. The outcome is that
we obtain the Salpeter correction with no explicit $\delta \mu$ appearance, 
with the stipulation that the chemical potential to be used is obtained from the number density
by doing the integral $n_e=\int d{\bf p} (2\pi)^{-3} f(E_p)$ and then solving for $\mu_e$
as a function of $n_e$. It appears to us that this gives the statement of results in the
best form for applications, and there is no explicit chemical potential shift in the result.

Thus we find reasons to question both of the above corrections; they do not contribute in the 
weakly coupled case, and we see no reason for them to be the dominant correction
in the strongly coupled case. That is not to say that the physical reasoning  behind
them is incorrect; other corrections are equally important. 

Turning to specific applications
we first focus our attention on ref. \cite{lang}, which is a comprehensive treatment of electron 
capture in the infall phase of a type II supernova event, with a full range of nuclear species, temperatures and densities. Here the screening corrections are a relatively minor
feature of the work; still they give effects ranging from a 25\% decrease in rates for a density
of $10^{10} gc^{-3}$, temperature $.75 MeV$ (with plasma parameter $\Gamma \approx 1$ ) to a factor of two at a density of $10^{12}$, $T=1.5 MeV, ~\Gamma\approx 3$ .
Our suggestion here is that these corrections be taken as an estimate of the magnitude  of the unknown error \underline{in either direction} due to plasma effects. If this turns out to create 
important uncertainties in the outcome of the collapse, then it will be time to worry.

In another venue, the accretion of matter onto the surface of a neutron star, the problems
could be much more serious. Here, at a depth where nuclear reactions stimulating transient
explosive phenomena can occur, it may be that electron capture on a particular
nuclear species is the stimulator and the rates will be extremely density and
temperature dependent \cite{kuul}, \cite{cooper}. In this environment, the plasma couplng $\Gamma$ will be much stronger than
those cited in the previous example, and we can expect the effects on rates to be larger,
and harder to estimate, even in magnitude. Furthermore, it is the temperature dependence
more than the absolute rate that rules in most explosion calculations. Since the plasma
corrections, whatever else, will be strongly temperature dependent, we could say that,
knowing nothing about them, we equally know nothing about the effective exponent 
in the temperature dependence of the rate.

There is a body of literature that presents unorthodox results on the effects 
of the plasma on reaction rates, even in the case of the weakly coupled solar
plasma. These results, if correct, would have significant impact on solar models. 
We wish it to be clear that the work of the present paper does not support any
of these claims.
One can find a critique in
ref. \cite{four} as well as a list of some of the articles in this category, though there are a more recent
examples that go down the same path \cite{dappen}. There are, in addition, a few
papers of the same genre addressing electron capture \cite{quarati1}, \cite{quarati2},
\cite{belyaev}. The last of these engendered a response in ref. \cite{davids}, one with which we
completely agree. But, that said, the reply in \cite{davids} was more in the vein of 
``We did a correct calculation and had no such terms; therefore your calculation
is wrong.", rather than tracking down specific mistakes. The same could be
said of the arguments given in ref. \cite{four}. Thus it may be worth adding
my own conclusion  about one thing that has gone wrong in at least some 
contributions of this genre. Let me characterize as an ``S-matrix approach" one in which
one starts the correction of a two body reaction rate by putting in some more bodies
from the plasma in the calculation, working
out the multi-body reaction rate in some perturbative way and then doing the 
thermal average (typically) over the momentum states of the plasma particles.
Looking at the actual mechanics in these papers, it appears to me that the
assumptions are tantamount to entirely leaving out Coulomb forces in the factor 
$\exp[-\beta H_a]$ in (\ref{M2}). It is a mistake to do this. We could put the conclusion
qualitatively as: A treatment strictly from a  multi-particle S-matrix, followed by 
averaging over thermal distributions, misses essential physics coming from the
fact that the initial particles sit in an interacting medium
\section{ Thinking about the strongly coupled and very degenerate case.}
\subsection{Ionic energy shift}
Returning
to the basic equation (\ref{M3}) that determines the kernel of the capture rate formula, we 
go back and think qualitatively about the ionic energy shift. We first state the case for the 
shift assumption, and then the counter-argument.
We take only electron kinetic energies and ionic potential energies in $H_{a,b}$, the electron 
potential energies not being relevent in the extremely degenerate system, at least not
to the effect that we are considering. Now 
we look 
first at the $H_b$ where there is a term $e_b \phi_I({\bf 0})$, the energy of ion $b$ in the fields of all of the other ions. 
If we forget about where $b$ came from, and when
it came into being,  and consider a problem in which it has been at ${\bf r}=0$ all along, we can in principle 
calculate the potential of the plasma ions at the position of $b$; it is of course a function of the charge $e_b=eZ_b$.  In classical simulations the energy shift is given 
by $\Delta E_b=T\,g(\Gamma_b)$ where $g(x)$ is of order $x$ and $\Gamma_b=Z^{5/3}_b e^2 3^{-1/3} (4 \pi n_e)^{1/3}$.
We can calculate the corresponding energy shift for initial ion $a$. From the time dependent exponentials 
in (\ref{M3}) we would infer that these corrections change the effective $Q$ value by an amount $\Delta E_a-\Delta E_b$.

But the actual term in the interaction, say, $H_a$, that induces an energy shift for ion $a$ is $e_a \phi ({\bf 0})$.  First, through this term, the plasma redistributes itself around the ion, creating a potential, which then shifts the energy of the ion through another action of $H_a$. In a perturbative calculation
the plasma is thereupon returned to its original state. If we let $H_a$ act only once, it will
transfer a bit of momentum from ion $a$ to the plasma. Of course the whole expression is a trace,
so everything has to be returned to its initial value. But that bit of momentum can be returned
to the plasma particles after the electron capture, through the interaction $H_b$. In the weakly
coupled case there is substantial cancellation between these categories of effects.
We cannot argue that in the strongly coupled case they would cancel in any particular way.
But we would not trust a result that takes one part and discards the other

\subsection{{\it Ab initio} numerical calculations?}
We take a box containing $N_I$ ions, not including $a$ or $b$, and $N_e$ electrons. The initial and final ions ${a,b}$ that participate in the capture process are at the point ${\bf r=0}$, and manifest themselves only in the potentials $H_a$ and $H_b$. 
In one respect 
we are at an advantage here, in comparison with a parallel treatment of the nuclear fusion problem, because
we probably do not care about regions in the space of the ions, $\bf R_i$, that are classically inaccessible, or nearly unaccessible at the temperature of the medium. Thus we can plausibly
take the ions to be absolutely classical, and entirely omit the ionic kinetic energy terms from $H_a$
and $H_b$.
 We take a basis
set of $N_e$ electrons at positions ${\bf r}_1,{\bf r_2}....{\bf r}_{N_e}$, and $N_I$ ions 
at positions ${\bf R_1},{\bf R}_2 ....{\bf R}_{N_I}$. Using the cyclic invariance of the trace in
(\ref{M3}) to move $\psi(0,0)$ from the last position to the first, and putting in the position basis
we have,
\vspace{.2 in}
\begin{eqnarray}
&M(t)=Z_P^{-1} \int d{\bf R}_1, .. d{\bf R}_{N_I},..d {\bf r}_2,..d{\bf r}_{N_{e}} 
 \int d {\bf r'}_2,..d{\bf r'}_{N_{e}}\times 
\nonumber\\  
 &\langle \{{\bf 0},{\bf r}_2,. {\bf r}_{N_e}\}, {\bf R}_1,.{\bf R}_{N_I} |e^{-(\beta-it)  H_a} |\{ {\bf 0},
 {\bf r'}_2,.{\bf r'}_{N_e}\}
{\bf R}_1,.{\bf R}_{N_I} \rangle 
 \nonumber\\
&  \times \langle \{{\bf r'}_2,.. {\bf r'}_{N_e}\}, {\bf R}_1,..{\bf R}_{N_I} |e^{-it  H_b} | 
 \{{\bf r}_2,..{\bf r}_{N_e}\},{\bf R}_1,..{\bf R}_{N_I} \rangle.
\nonumber\\
\label{basic2}
\end{eqnarray}
We have here shifted from the grand canonical ensemble to the canonical ensemble, with a
definite number of electrons. The \{ \} around the electron basis states signify a total 
anti-symmetrization. The ionic part requires no symmetrization.  Note that the intermediate integration
is over electronic positions only. We have taken the classical limit for the ions, in which the ionic kinetic energy
term is omitted and the ionic coordinates within $H_a$ and $H_b$
remain at their external values.

Could we in principle use he trick of ref. \cite{militzer},
  $ \exp(-\beta H)=[\exp(-\beta H /N_S)]^{N{_S}}$  taking
 $N_S$ is sufficiently large so that we can make a lowest order perturbative expansion in the 
 Coulomb potentials in each individual factor? Each factor would then be at a sufficiently high temperature, $N_ST$, to validate this lowest order expansion. This would be at best an expensive
calculation because $N_1$, $N_e$ and $N_I$ all have to be fairly large, and between each of the $N_1$ factors in the product one must integrate over the full manifold of some ${\bf r}_1'...{\bf r}'_{N_e}$, not to mention the enormous anti-symmetrizations required. And the final kiss of death would be:
because of the real time exponentials, we would have rapidly oscillating integrands.

But perhaps instead, we could start by taking a very degenerate plasma. The electron sea is so stiff that
it does nothing, except that one gets captured, creating an electron hole in the final state. Should
we be worried about the effects of the (ionically) irregular plasma on what that hole does?
Perhaps somewhat symbolically, we can write a much simpler equation,

\begin{eqnarray}
&M(t)=Z_P^{-1} \int d{\bf R}_1, .. d{\bf R}_{N_I} 
 \langle  {\bf R}_1,..{\bf R}_{N_I} |e^{-(\beta-it)  H_a} |
 {\bf R}_1,.{\bf R}_{N_I} \rangle 
 \nonumber\\
 & \times \langle {{\bf \bar r}={\bf 0}, \bf R}_1,..{\bf R}_{N_I} |e^{-it  H_b} | 
 {\bf \bar r}={\bf 0}, {\bf R}_1,..{\bf R}_{N_I} \rangle.
\nonumber\\
\,
\label{basic3}
\end{eqnarray}
Is it possible to calculate the matrix elements for the hole to go from the origin back to the origin in time $t$ under the
action of $H_b$, for a big set of irregular distributions of ions ? If one had to know
the answer; the eikonal method might serve.

This work was supported 
in part by NSF grant PHY-0455918. 

\section{Appendix. Perturbation calculations}

To demonstrate the methods we divide the respective Hamiltonians, as $H_{a,b}=H_0+H_{a,b}^I$ , in which the interaction term contains all Coulomb  interactions of the electrons in the plasma and the interactions of the distinguished ions $I_a$ and $I_b$ with both electrons and plasma ions. Coulomb interactions among the ions in the plasma are
retained in $H_0$. \footnote{In our terminology, the distinguished ions $I_a$ and $I_b$ are not part
of the ``plasma". Since Fermi statistics prevents
us from distinguishing an electron; we include all electrons in the ``plasma".}
We denote the charges of the electron and two ions
respectively as $e,e_a,e_b$, where $e_b=e_a-1$ obtaining,

\begin{eqnarray}
&H_a^I=\int d{\bf r}\Bigr[-e [\phi_I({\bf r})+{1 \over 2}\phi_e({\bf r})]n_e({\bf r})+e_I n_I ({\bf r})
\phi({\bf r})\Bigr] 
\nonumber\\
&+e_a \phi({\bf 0}),
\label{hia}
\end{eqnarray}
and
\begin{eqnarray}
&H_b^I=\int d{\bf r}\Bigr [-e [\phi_I({\bf r})+{1 \over 2}\phi_e({\bf r})]n_e({\bf r})+
e_I n_I ({\bf r})\phi({\bf r})\Bigr] 
\nonumber\\
&+e_b \phi({\bf 0}).
\label{hib}
\end{eqnarray}

Here $\phi_e$, $\phi_I$ are the respective electric potentials produced by the electrons and $\phi=\phi_I+\phi_e$
is the total potential of the plasma. We have not included the basic Coulomb interactions between
electrons and $I_a$ and $I_b$. But if we repeated the calculations to come using
Coulomb wave-functions instead of plane waves we would not find any medium-dependent terms that are of leading order, $e^2\kappa_D$. There is of course a conventional Coulomb correction in the
absence of the plasma; perturbatively it begins at order $e^2$, and for solar core capture
in $^7$Be is larger than the order $e^2 \kappa_D$ screening corrections. At the level of the
present paper this vacuum contribution can just be added to the answer; Coulomb wave-functions are
not needed for calculating the rest of the corrections.

We introduce the interaction picture through the identities,

\begin{eqnarray}
e^{-H_a (\beta-it)}=e^{-H_0(\beta-i t)} \Omega_+(-i\beta, t)\,,
\label{int1}
\end{eqnarray}
and,
\begin{eqnarray}
e^{-i H_b t}= \Omega_-(0,t) e^{-i H_0 t },
\label{int2}
\end{eqnarray}
where,
\begin{eqnarray}
\Omega_a^{(+)}(-i\beta, t)=\exp\Bigr [i \int_{-i \beta}^t dt' \hat H_a^I(t')\Bigr ]_+\,,
\label{om1}
\end{eqnarray}
and
\begin{eqnarray}
\hat H^I_a(t)=e^{i H_0 (t+i\beta)} H^I_a e^{-i H_0 (t+i\beta)}.
\end{eqnarray}
In (\ref{om1}) the subscript, $+$, stands for time ordering of the integrals.
Similarly we have,
\begin{eqnarray}
 \Omega_b^{(-)}(0,t)=\exp\Bigr [-i \int_{0}^t dt' \hat H_b^I(t')\Bigr ]_- 
\label{om2}
\end{eqnarray}
where
\begin{eqnarray}
\hat H^I_b(t)=e^{i H_0 t} H^I_b e^{-i H_0 t},
\end{eqnarray}
and now the integral is anti-time-ordered.
Substituting in (\ref{M2}) we obtain
\begin{eqnarray}
&M(t)=Z_P^{-1}{\rm Tr} \Bigr [e^{-H_a  (\beta-i t)}\psi_e^\dagger({\bf 0},0)   e^{-i H_b t}\psi_e({\bf 0},0)\Bigr ]=
\nonumber\\
&{Z_P^{-1}\rm Tr} \Bigr [e^{-H_0(\beta -i t)}\Omega^{(+)}_a(-i \beta,t)\psi_e^\dagger({\bf 0},0) 
\Omega^{(-)}_b (0,t) e^{-i H_0 t}
\nonumber\\
&\times \psi_e({\bf 0},0)\Bigr ]
={Z_P^{-1}\rm Tr} \Bigr [e^{-H_0 \beta }\Omega^{(+)}_a(-i \beta,t)\psi_e^\dagger({\bf 0},0) 
\nonumber\\
&\times\Omega^{(-)}_b (0,t) e^{-i H_0 t}\psi_e({\bf 0},0)e^{i H_0 t}\Bigr ].
\nonumber\\
\label{ans}
\end{eqnarray}
We expand the $\Omega$ factors in powers of the coupling, retaining only terms of order $e^2$, (or $e_a^2, e e_a, e_a^2, etc)$. Schematically, each will have powers of $e$ coming from the thermal expectation of a product of two
ion-electric-potential $\phi({\bf r})$ operators; these are determined by the Hamiltonian $H_0$. If this potential-potential
correlator is itself expanded in powers of $e$, the expansion begins with terms of order $e^2$, so that the
rate corrections would be of order $e^4$ but the terms have an infrared singularity from the long range Coulomb force,
and when that is properly regulated the entire correction to the rate will be of order $e^3$. In the present paper we
pursue only these $e^3$ terms, the neglected terms will go as power $e^4$ or higher. For this purpose we shall need the part of the field correllation function that is the most singular as $\kappa_D$ goes to zero,
\begin{eqnarray}
\Bigr \langle \phi ({\bf k},t') \phi ({\bf -k'},t'') \Bigr \rangle_{IR}\approx 4 \pi \delta ({\bf k-k'})
\beta^{-1}\kappa_D^2{1\over (k^2+\kappa_D^2)k^2},
\nonumber\\
\,
\label{corr}
\end{eqnarray}
which is time independent.

We now expand $\Omega_a^{(+)}$
\begin{eqnarray}
\Omega _a^{(+)}(-i\beta,t)=1-i \int_{-i\beta}^t dt_1\hat H_a^I(t_1)
\nonumber\\
-\int_{-i\beta}^t dt_1 \hat H_a^I(t_1)\int_{-i\beta}^{t_1}dt_2 \hat H_a^I(t_2),
\nonumber\\
\,
\label{oma}
\end{eqnarray}
and
$\Omega_b^{(-)}$,
\begin{eqnarray}
\Omega _b^{(-)}(0,t)=1+i \int_{0}^t dt_1\hat H_b^I(t_1)
\nonumber\\
-\int_{0}^t dt_1\hat H_b^I(t_1)\int_{0}^{t_1}dt_2 \hat H_b^I(t_2).
\nonumber\\
\,
\label{omb}
\end{eqnarray}
Before getting bogged down in the evaluation of a multiplicity of terms we think for a moment about the general
form of the expressions generated by the expansion. Leaving out most of the arguments of the functions and not concerning ourselves
with orders of factors, we note that the form is the expectation value,
\begin{eqnarray}
\Bigr \langle \Bigr [-e\int n_e [\phi_I+{1 \over 2} \phi_e]+e_{\{a,b\}}\phi(0)\Bigr ]
\nonumber\\
\Bigr [-e\int n_e [\phi_I+{1 \over 2} \phi_e]+e_{\{a,b\}}\phi(0)\Bigr ]\psi^\dagger_e\psi_e \Bigr \rangle.
\label{horrible}
\end{eqnarray}

Now we state some results of peering into the term by term outcome:

1.) The electron operators within $\psi_e, \psi_e^\dagger$ each get paired with an electron operator in
an $n_e$ or $\phi_e$ term earlier in the expression. The terms in which they eat each other are ``disconnected",
i.e. their contribution gets canceled by a perturbation in the partition function in the denominator of (\ref{ans}),
in conjunction with an unperturbed numerator.
 
2.) Terms where both $\psi$ and $\psi^\dagger$ pair with the operators from the same big square bracket in (\ref{horrible}) are not infrared divergent in the absence of screening and will therefore not contribute in leading order.

3.) Bearing in mind 2.) above, when a single $\psi$ or $\psi^\dagger$ is to be paired with an operator
from the product, 
\begin{eqnarray}
{1 \over 2} \int d{\bf r }\phi_e ({\bf r}) n_e({\bf r})=\int d{\bf r}\int d {\bf r'} |{\bf r-r'}|^{-1} n_e({\bf r}) n_e({\bf r'}),
\nonumber\\
\,
\end{eqnarray}
then the pairing will be, in effect, the same when it is with an operator in $n_e$ or when it is with one in $\phi_e$.
Thus if we add a stipulation to the calculation of (\ref{horrible}) that the explicit  $\psi_e$ and $\psi^\dagger$
are never to be paired with
$\phi_e$ in this kind of term, we can compensate
by multiplying the $\phi_e$'s in (\ref{horrible}) by 2. To put it more simply, we can replace $H_{a,b}$ of  (\ref{hia})
and (\ref{hib}) by,
\begin{eqnarray}
H_a^I=\int d{\bf r} [-e \phi({\bf r})n_e({\bf r})]+e_a \phi({\bf 0}),
\nonumber\\
H_b^I=\int d{\bf r} [-e \phi({\bf r})n_e({\bf r})]+e_b \phi({\bf 0}),
\label{HI}
\end{eqnarray}
with the above stipulation on computation.
These operators $\hat H_{a,b}^I(t)$ in the interaction picture are now given in terms of the electron creation and 
annihilation operators $a^\dagger(p),a(p)$ by,
\begin{eqnarray}
\hat H_{a,b}^I(t)=\int {d^3 k\over (2 \pi)^3}\phi (k,\tau)
\nonumber\\
\times\Bigr [e_{a,b} 
-e\int {d^3 p\over(2 \pi)^3} a_{\bf p+k}^\dagger a_{\bf p}e^{i(E_{p+k}-E_p)\tau}\Bigr],
\label{hint}
\end{eqnarray}
where $\tau=t+i\beta$ for the incoming system $a$ and $\tau=t$ for the final system.
In the the small ${\bf k}$ limit we can set $\tau=0$ in the final bracket in (\ref{hint}). 
We do
keep an infinitesimal ${\bf k}$ in the indices of the operators $a_{\bf p+k}$, as we shall see below. Since the infrared part of the correlator is also time independent, there is in effect no
time dependence. 
Putting (\ref{hint}) into (\ref{oma}) and (\ref{omb}), and then putting the results into (\ref{ans})
and using (\ref{corr})
we obtain,
\begin{eqnarray}
M(t)=M_0(t) -4 \pi \int {d{\bf k}\over (2 \pi)^3}{\beta^{-1} \kappa_D^2 \over k^2 (k^2+\kappa_D^2)}
\nonumber\\
\times \Bigr [
({t^2\over 2} +i \beta t -{\beta^2\over 2})M_1(t)-(t+i\beta)t M_2(t)+{t^2\over 2} M_3(t)\Bigr ],
\nonumber\\
\label{M(t)}
\end{eqnarray}
where
\begin{eqnarray}
M_1(t)={\rm Lim}_{k \rightarrow 0}\Bigr \langle \Bigr [e_a-e\int d{\bf p_3}\,a_{{\bf p}_3+{\bf k}}^\dagger
a_{{\bf p}_3}\Bigr]
\nonumber\\
\times \Bigr[e_a -e \int d{\bf p_4}\,a_{{\bf p}_4-{\bf k}}^\dagger
a_{{\bf p}_4}\Bigr]\int d{\bf p}_1d{\bf p} e^{i E_pt} a_{{\bf p}_1}^\dagger a_{{\bf p}}\Bigr \rangle,
\label{Ma1}
\end{eqnarray}
\begin{eqnarray}
M_2(t)={\rm Lim}_{k \rightarrow 0}\Bigr \langle \Bigr [e_a-e\int d{\bf p_3}\,a_{{\bf p}_3+{\bf k}}^\dagger
a_{{\bf p}_3}\Bigr]
\nonumber\\
\times \int d{\bf p}_1a_{{\bf p}_1}^\dagger \Bigr[e_b-e \int d{\bf p_4}\,a_{{\bf p}_4-{\bf k}}^\dagger
a_{{\bf p}_4}\Bigr]\int d{\bf p} e^{i E_pt} a_{{\bf p}}\Bigr \rangle,~~
\end{eqnarray}
\begin{eqnarray}
&M_3(t)={\rm Lim}_{k \rightarrow 0}\Bigr \langle \int d{\bf p}_1a_{{\bf p}_1}^\dagger \Bigr [e_b-e\int d{\bf p_3}\,a_{{\bf p}_3+{\bf k}}^\dagger
a_{{\bf p}_3}\Bigr] 
\nonumber\\
&\times \Bigr[e_b-e \int d{\bf p_4}\,a_{{\bf p}_4-{\bf k}}^\dagger
a_{{\bf p}_4}\Bigr]\int d{\bf p} e^{i E_pt} a_{{\bf p}}\Bigr \rangle,
\label{Ma3}
\end{eqnarray}
and $M_0$ is the unperturbed part,
\begin{eqnarray}
M_0(t)= \int d{\bf p}_1\,d{\bf p}\Bigr \langle a_{{\bf p}_1}^\dagger a_{\bf p} e^{i E_{\bf p}t} \Bigr \rangle
= \int {d^3p }f(p)e^{iE_pt}.
\nonumber\\
\,
\label{m0}
\end{eqnarray}
Here $f(p)$ the Fermi distribution, $f(E_p)=(1+e^{\beta(E_p-\mu_e)})^{-1}$ and the electron chemical potential is
determined from the electron density \underline{to zeroth order in $e$} by $n_e=\int d{\bf p}(2\pi)^{-3} f(p)$. When we calculate perturbative 
corrections we must explicitly include a correction that readjusts the relation between chemical potential and
density, which we shall do below.

The thermal averages in ({\ref{Ma1})-(\ref{Ma3}) are readily calculated.  A number of terms are disconnected
and canceled by perturbative corrections to the partition function: a) all terms of order $e^2$ in which
$p=p_1$, i.e. in which operators in $\psi({\bf 0})$ are paired with those in $\psi^\dagger ({\bf 0})$; 
b) some more that are discussed below.
The rest of the terms can be expressed in terms of $M_0$  of (\ref{m0}), the function $M_A$,

\begin{eqnarray}
M_A(t)= \int {d^3 p \over (2 \pi)^3}[f(p)]^2 e^{iE_pt},
\label{mA}
\end{eqnarray}

and a time independent constant, $R$,
\begin{eqnarray}
M_1(t)=M_2(t)=M_3(t)=~~~~~~~~~~~~~~
\nonumber\\
(e_a-e)^2 M_0(t)+(2 e_a  e-e^2) M_A(t)+ e^2RM_0(t)\,,
\label{ms}
\end{eqnarray}
where $R$ is a time independent constant that is proportional to the system volume and that will be canceled
by a corresponding perturbation of the partition function.

We reconstitute the complete time dependent function $M(t)$ which when inserted in (\ref{orig2}) gives the rate.
\begin{eqnarray}
M(t)=\Bigr[1+{1\over 2}(e_a-e)^2 \kappa_D \beta\Bigr ]M_0(t)
\nonumber\\
+{1\over 2}(2e e_a-e^2)  \kappa_D \beta M_A(t)
-{1\over 2 }\kappa_D \beta e^2 RM_0(t)].
\label{ans3}
\end{eqnarray}
Now we calculate the changes of the same order to the partition function, where the system is governed by $H_a$,
\begin{eqnarray}
&Z_{P}=\Bigr \langle e^{-\beta H_0} \Omega_a^{+} (-i\beta, 0)\Bigr  \rangle_\beta= 
\Bigr \langle e^{-\beta H_0} \Bigr [1+ 
\nonumber\\
 &\int_{-i \beta} ^0 dt_1\int_{-i \beta}^{t_2} dt_2 
\times [e_a \phi({\bf 0},t_1)+e\int d{\bf r}  \phi ({\bf r}, t_1) n_e ({\bf r}, t_1)] 
\nonumber\\
&\times [e_a \phi({\bf 0},t_2)+
e\int d{\bf r'}  \phi ({\bf r'}, t_2) n_e ({\bf r'}, t_2)]\Bigr ]\Bigr \rangle_\beta.
\nonumber\\
\label{part}
\end{eqnarray}

The calculation follows the lines outlined above, but it is simpler. We obtain a contribution of
order $e_a^2$ that is the same as the order $e_a^2$ term in (\ref{ans3}), but without the 
factor of $M_0(t)$. Therefore in conjunction with the $1 \times M_0(t)$ term in (\ref{ans}),
we see that the partition function perturbation removes the $e_a^2 \kappa_D \beta M_0(t)$
term from (\ref{ans3}). The other non-vanishing term coming from (\ref{part}) is of order $e^2$
and  it exactly removes the final term, proportional to $R$, from (\ref{ans3}).

Next we must recognize that in our calculation
the chemical potentials were held fixed while we calculated perturbations to the rates. But the relation
between number density and chemical potential has corrections of exactly the same order
as the rates. In the next section of the appendix we calculate the changes in this relation, and we find that the re-expression
of the results in terms of the corrected densities removes the remaining terms of order $e^2$ 
from (\ref{ans3})\footnote{ This mirrors the situation in the
case of a perturbative calculation of plasma corrections ot fusion rates, as discussed in ref. \cite{bs}},
leaving the final result for the corrections of
leading order as simply, 

\begin{eqnarray}
M(t)=M_0(t)-e \,e_a  \kappa_D \beta\Bigr (M_0(t)-M_A(t)\Bigr ).
\label{ans7}
\end{eqnarray}

\subsection{Keeping $n_e$ fixed}
In sec. 4 we used a relation between the change in the electron number density, $\delta n_e$, when we turn on 
Coulomb interactions in the medium keeping the chemical potential fixed, and the chemical potential change $\delta \mu_e$
that is then required to restore the original electron density.  Then this  $\delta \mu_e$, inserted as a correction into
the Fermi distribution in the unperturbed rate function $F_0(t)$ generates a change that must be added to the terms shown
in (\ref{ans3}), in order that the complete result give the change in rate induced by the Coulomb forces
while keeping the number density, rather than the chemical potential, constant. We have 
\begin{eqnarray}
 n_e=Z^{-1}Tr[e^{\beta H } \psi^\dagger ({\bf 0})\psi({\bf 0})]
\nonumber\\
=Z^{-1}Tr[e^{\beta H_0 } \Omega_e^+(0,-i\beta )\psi^\dagger ({\bf 0})\psi({\bf 0})],
\end{eqnarray}
where the operator $\Omega_e^+(0,-i\beta )$ is expanded in exactly the form shown in (\ref{oma}), except that
we omit the interaction of the ion with the plasma in the calculation; that is we take $e_a=e_b=0$. Only the second order in the expansion
contributes, and we see that the answer is given in terms functions that we have already exhibited, now evaluated at
$t=0$,
\begin{equation}
n_e=n_e^{(0)}+{1\over 2}e^2 \kappa_D \beta [M_0(0)-M_A(0)],
\end{equation}
This gives a density shift that is compensated by a chemical potential shift,
\begin{eqnarray}
\delta \mu_e={1\over 2} e^2 \kappa_D,
\end{eqnarray}

We must compensate with a shift $\delta \mu_e$ in the function $M_0(t)$. Expanding, this gives a shift in the 
function that determines the rate,
\begin{eqnarray}
\delta_e  M(t)=-{1\over 2}e^2 \beta \kappa_D [M_0(t)-M_A(t)].
\end{eqnarray}

  \end{document}